\RequirePackage{arydshln}
\documentclass[aps,twocolumn,nofootinbib,superscriptaddress,preprintnumbers,pra,10pt,floatfix]{revtex4-1}

\usepackage[utf8]{inputenc}

\usepackage{float}
\usepackage{amsmath,amssymb}
\usepackage{dsfont} 
\usepackage{hyperref}
\usepackage{graphicx}
\usepackage{enumitem}
\usepackage{mathtools}
\usepackage{bbold}
\usepackage{multirow}
\usepackage{ytableau}
\usepackage{youngtab}
\usepackage{braket}
\usepackage{soul}
\usepackage{cancel}
\usepackage{xcolor}
\usepackage[normalem]{ulem}
\usepackage[export]{adjustbox}

\usepackage{nicematrix}

\usepackage{fontawesome}

\usepackage{bbm}
\usepackage{mathrsfs}
\usepackage{bm}

\usepackage{slashed}

\usepackage{lipsum}

\usepackage{colortbl} 
\definecolor{Gray}{gray}{0.95}
\definecolor{RGray}{gray}{0.290}
\definecolor{CGray}{gray}{0.92}

\usepackage{arydshln}

\definecolor{blueLinks}{RGB}{0,84,159}
\newcommand{\myColor}[0]{blueLinks}
\hypersetup{
	colorlinks, 
	bookmarksopen, 
	bookmarksnumbered,
	pdftoolbar=false, 
	pdfmenubar=false, 
	pdffitwindow=false, 
	pdfstartview={FitH},	 
	citecolor=\myColor, 
	linkcolor=\myColor,	
	urlcolor=\myColor, 
	pdftitle={On the EFT validity for Drell-Yan tails at the LHC}, 
	pdfsubject={EFT validity - Drell-Yan}, 
	pdfnewwindow=true, 
	linktoc=all
}

\makeatletter
\g@addto@macro\bfseries{\boldmath}
\makeatother

\makeatletter
\renewcommand\paragraph{\@startsection{paragraph}{4}{\z@}%
                                    {3.25ex \@plus1ex \@minus.2ex}%
                                    {-1em}%
                                    {\normalfont\normalsize\bfseries}}
\makeatother

\allowdisplaybreaks
\begin{document}


\title{Neutrinoless Double-Beta Decays from Operator Mixing}
\author{J.~de Vries}
\email{j.devries4@uva.nl}
\affiliation{Institute for Theoretical Physics, University of Amsterdam, 
Science Park 904, 1098 XH Amsterdam, The Netherlands}
\affiliation{Theory Group, Nikhef,
Science Park 105, 1098 XG, Amsterdam, The Netherlands}
\author{S.~Fajfer}
\email{svjetlana.fajfer@ijs.si}
\affiliation{Jožef Stefan Institute, Jamova 39, P. O. Box 3000, 1001 Ljubljana, Slovenia}
\author{L.P.S~Leal}
\email{luighi.leal@usp.br}
\affiliation{Instituto de Física, Universidade de São Paulo, São Paulo – São Paulo 05580-090, Brazil.}
\author{O.~Sumensari}
\email{olcyr.sumensari@ijclab.in2p3.fr}
\affiliation{IJCLab, P\^ole Th\'eorie (Bat.~210), CNRS/IN2P3 et Universit\'e Paris-Saclay, 91405 Orsay, France}
\author{R.~Zukanovich Funchal}
\email{zukanov@if.usp.br}
\affiliation{Instituto de Física, Universidade de São Paulo, São Paulo – São Paulo 05580-090, Brazil.}
%

\begin{abstract}
\vspace{5mm}
We perform a complete one-loop Renormalization Group (RG) analysis of neutrinoless double-beta decays within the Standard Model Effective Field Theory (SMEFT). Although several effective operators do not contribute to these processes at tree level, we show that they can generate sizable contributions through operator mixing. By accounting for the full one-loop RG evolution within the SMEFT, we find that the resulting constraints provide the most stringent bounds on numerous dimension-seven operators, improving upon limits derived from meson decays. We also highlight the importance of contributions beyond the first leading logarithm, which can provide the leading effects for specific operators and flavor combinations.
\vspace{3mm}
\end{abstract}

\maketitle

\allowdisplaybreaks

\newcommand{\cmd}[1]{{\texttt{#1}}\xspace}
\newcommand{\dd}{\mathop{}\!\mathrm{d}}
\newcommand{\braces}[1]{\left\lbrace #1 \right\rbrace}
\newcommand{\brackets}[1]{\left( #1 \right)}
\newcommand{\squarebrackets}[1]{\left[ #1 \right]} 
\newcommand{\angles}[1]{\left\langle #1\right\rangle}
\newcommand{\abs}[1]{\left\lvert #1 \right\rvert}
\newcommand{\norm}[1]{\left\Vert #1 \right\Vert}
\renewcommand{\Im}{\mathop{\mathrm{Im}}}
\renewcommand{\Re}{\mathop{\mathrm{Re}}}
\renewcommand{\L}{\mathcal{L}}
\newcommand{\ord}[1]{\mathcal{O}\!\left( #1 \right)}
\newcommand{\ineqgraphics}[1]{\vcenter{\hbox{\includegraphics[]{#1}}}} 

\newcommand{\SMEFT}[0]{\text{SM\,EFT}\xspace}
\newcommand{\vev}[0]{vacuum expectation value\xspace}
\newcommand{\QFT}[0]{quantum field theory\xspace}
\newcommand{\SU}[1]{\mathrm{SU}(#1)}
\newcommand{\C}[2]{C_{\underset{#2}{#1}}}
\newcommand{\vT}[0]{\overline{v}_T}

\newcommand{\cO}{{\mathcal O}}
\newcommand{\cQ}{{\mathcal Q}}
\newcommand{\cC}{{\mathcal C}}
\newcommand{\cL}{{\mathcal L}}
\newcommand{\cH}{{\mathcal H}}
\newcommand{\cI}{{\mathcal I}}
\newcommand{\cA}{{\mathcal A}}
\newcommand{\cB}{{\mathcal B}}
\newcommand{\cF}{{\mathcal F}}
\newcommand{\cP}{{\mathcal P}}
\newcommand{\cS}{{\mathcal S}}
\newcommand{\cT}{{\mathcal T}}
\newcommand{\cU}{{\mathcal U}}
\newcommand{\cN}{{\mathcal N}}
\newcommand{\cM}{{\mathcal M}}
\newcommand{\cZ}{{\mathcal Z}}

\newcommand{\SMrep}[3]{({\bf #1},\,{\bf #2},\,#3)}
\newcommand{\SMrepbar}[3]{({\bf \bar#1},\,{\bf #2},\,#3)}

\definecolor{myGreen}{rgb}{0.0, 0.7, 0.1}

\newcommand{\dario}[1]{{\color{blue} [OS:#1]}}
\newcommand{\olcyr}[1]{{\color{orange} [OS:#1]}}
\newcommand{\felix}[1]{{\color{myGreen} [FW: #1]}}
\newcommand{\felixx}[1]{{\color{myGreen} #1}}
\newcommand{\lukas}[1]{{\color{purple}[L.A.: #1]}}
\newcommand{\lukkas}[1]{{\color{purple} #1}}
\newcommand{\matheus}[1]{{\color{blue} [MM:#1]}}



\section{Introduction}

The discovery of neutrino oscillation has established beyond doubt that neutrinos are massive, providing clear evidence for physics beyond the Standard Model (SM). Over the past several decades, experiments based on solar, atmospheric, and reactor neutrinos have made substantial progress in determining the neutrino squared-mass differences and mixing parameters~\cite{SajjadAthar:2021prg}. However, several questions remain unanswered in the neutrino sector~\cite{Huber:2022lpm}. The most fundamental of these are the specific nature and the ultraviolet mechanism that gives rise to the tiny neutrino masses.

The most compelling assumption, based on Effective Field Theories (EFTs), is that neutrino masses are a manifestation of the only dimension-five operator allowed by the SM gauge symmetry~\cite{Weinberg:1979sa},
\begin{equation}
    \label{eq:Weinberg-operator}
    \mathcal{O}^{(5)}_{LH} = \epsilon_{ab} \epsilon_{mn} \big{(}L_a^T C L_m \big{)} H_b H_n\,,
\end{equation}

\noindent where $H$ denotes the SM Higgs doublet,  $L$ is the lepton doublet  and $C$ stands for the charge-conjugation matrix. $SU(2)_L$ indices are denoted by Latin letters, and flavor indices are omitted, for simplicity, in the above equation. This operator can explain the smallness of neutrino masses by the $v/\Lambda$ suppression, where $\Lambda \gg v$ is the EFT cutoff scale and $v$ denotes the Higgs vacuum-expectation value. The primary implication of such an operator beyond neutrino masses is that it induces Lepton Number Violating (LNV) processes, such as neutrinoless double-beta decays ($0\nu\beta\beta$), which are strictly forbidden in the SM. Several experiments have set strong limits on the $0\nu\beta\beta$ half-life in recent years~\cite{KamLAND-Zen:2024eml,GERDA:2020xhi,Majorana:2022udl,EXO-200:2019rkq,CUORE:2022piu}, which will be significantly improved by the next-generation experiments, reaching a sensitivity of $10^{27}$–$10^{28}$~years in the near future~\cite{LEGEND:2021bnm,nEXO:2021ujk,SNO:2021xpa}.

EFT analyses of $0\nu\beta\beta$ decays have been performed in several studies~\cite{Cirigliano:2017djv,Cirigliano:2018yza,Fridell:2023rtr}, also  including Renormalization Group (RG) evolution effects for a subset of LNV operators~\cite{Graf:2025cfk}. While the naive EFT power counting suggests that contributions from higher-dimensional operators to these processes are suppressed in comparison to the Weinberg operator defined in Eq.~\eqref{eq:Weinberg-operator}, important caveats arise once the EFT is matched onto concrete ultraviolet scenarios. Notably, the dimension-five effective coefficient can be loop-suppressed with respect to the dimension-seven ones. Furthermore, the latter can induce chirality-enhanced contributions to $0\nu\beta\beta$ decays ($\propto E/m_\nu$), where $E\sim 100~\mathrm{MeV}$ is the typical energy of these nuclear processes. These features are combined within radiative models for neutrino masses, where dimension-seven operators can induce the dominant contributions to $0\nu\beta\beta$ decays, as demonstrated, e.g., in scenarios involving scalar leptoquarks~\cite{Hirsch:1996ye,Fajfer:2024uut,Fridell:2024pmw}.

\begin{table*}[!t]
    \renewcommand{\arraystretch}{2.1}
    \centering
        \centering
        \begin{NiceTabular}{c|cc||c|cc}
        \hline
        \multicolumn{6}{c}{ $d=7$ operators}  \\ \hline\hline
        $\psi^2 H^4$   & $\mathcal{O}_{LH}^{(7)}$ & $\epsilon_{ab}\,\epsilon_{mn} \big{(}L_a^T C L_m \big{)} H_b H_n \big{(}H^\dagger H\big{)}$&  \multirow{5}{*}{$\psi^4 H$} & \cellcolor[gray]{.92}$\boldsymbol{\mathcal{O}^{(7)}_{\bar{d}LQLH1}}$& $\cellcolor[gray]{.92}\epsilon_{ab}\,\epsilon_{mn} \big{(}\bar{d}L_a\big{)}\big{(}Q_b^T C L_m\big{)}H_n$\\ \cline{1-3}
        $\psi^2 H^3 D$ & \cellcolor[gray]{.92} $\boldsymbol{\mathcal{O}^{(7)}_{LeHD}}$& \cellcolor[gray]{.92} $\epsilon_{ab}\,\epsilon_{mn}\big{(}L_a^T C \gamma_\mu e \big{)}H_b H_m \big{(}iD^\mu H\big{)}_n$ & & $\cellcolor[gray]{.92}\boldsymbol{\mathcal{O}^{(7)}_{\bar{d}LQLH2}}$& $\cellcolor[gray]{.92}\epsilon_{am}\,\epsilon_{bn} \big{(}\bar{d}L_a\big{)}\big{(}Q_b^T C L_m\big{)}H_n$ \\\cline{1-3}
        \multirow{2}{*}
        {$\psi^2 H^2 D^2$} & \cellcolor[gray]{.92} $\mathcal{O}_{LHD1}^{(7)}$ & \cellcolor[gray]{.92}$\epsilon_{ab}\epsilon_{mn}\big{(}L_a^T C (D_\mu L)_b\big{)} H_m \big{(}D^\mu H\big{)}_n$ &  & $\cellcolor[gray]{.92}\boldsymbol{\mathcal{O}^{(7)}_{\bar{Q}uLLH }}$ & $\cellcolor[gray]{.92}\epsilon_{ab}\big{(} \overline{Q}_m u\big{)}\big{(} L_m^T C L_a\big{)}H_b$\\
         & \cellcolor[gray]{.92} $\mathcal{O}_{LHD2}^{(7)}$ & \cellcolor[gray]{.92} $\epsilon_{am}\epsilon_{bn}\big{(}L_a^T C (D_\mu L)_b\big{)} H_m \big{(}D^\mu H\big{)}_n$ & & $\cellcolor[gray]{.92}\boldsymbol{\mathcal{O}^{(7)}_{\bar{d}LueH}}$ & $\cellcolor[gray]{.92}\epsilon_{ab} \,\big{(}\bar{d} L_a \big{)}\big{(}u^T C e\big{)} H_b$\\ \cline{1-3}
        \multirow{2}{*}{$\psi^2 H^2 X$} & $\mathcal{O}_{LHB}^{(7)}$ & $\epsilon_{ab}\,\epsilon_{mn} \,\big{(}L_a^T C \sigma^{\mu\nu} L_m\big{)} H_b H_n B_{\mu\nu}$ &  & $\mathcal{O}_{\bar{e}LLLH}^{(7)}$& $\epsilon_{ab}\epsilon_{mn} \big{(}\bar{e} L_a\big{)}\big{(}L_b^T C L_m\big{)}H_n$\\ \cline{4-6}
        & \cellcolor[gray]{.92} $\mathcal{O}_{LHW}^{(7)}$ & \cellcolor[gray]{.92}$\epsilon_{ab} \,\big{(}\epsilon \tau^I\big{)}_{mn} \,\big{(}L_a^T C \sigma_{\mu\nu} L_m\big{)} H_b H_n W_{\mu\nu}^I$ & $\psi^4 D$ & \cellcolor[gray]{.92}$\mathcal{O}^{(7)}_{\bar{d}uLLD}$& \cellcolor[gray]{.92}$\epsilon_{ab} \big{(}\bar{d}\gamma_\mu u\big{)}\big{(}L_a^T C (iD^\mu L)_b\big{)}$\\
        \end{NiceTabular}
    \caption{\small \sl Dimension-seven LNV operators in the SMEFT in the basis of Ref.~\cite{Zhang:2023kvw}. The Pauli matrices are denoted by $\tau^I$ (with $I=1,2,3$), $\epsilon=i\tau^2$ is the Levi-Civita tensor and $C=i\gamma_2\gamma_0$ is the charge conjugation matrix. $SU(2)_L$ are denoted by Latin letters and, for simplicity, flavor indices are omitted. The operators that contribute at tree level to $0\nu\beta\beta$ decays for first-generation fermions are highlighted in gray. Among these operators, we distinguish with boldface those with contributions that are not suppressed by light fermion masses, see~Appendix~\ref{app:matching}.}
    \label{tab:EFT-LNV-d7}
\end{table*}

In this paper, we will systematically investigate the impact of RG-induced mixing on EFT analyses of $0\nu\beta\beta$ decays involving dimension-seven operators, extending previous studies that have focused either on the diagonal RG-evolution or on specific operators. We will consider the full set of Standard Model Effective Field Theory (SMEFT)  operators, with a general quark flavor structure, and the complete one-loop anomalous dimension matrix computed in Ref.~\cite{Zhang:2023kvw}. These mixing effects can be particularly relevant for operators that do not contribute at tree level to these processes, with the leading contributions appearing only at loop level. This is the case for operators with heavy quark flavors, which mix into those involving valence quarks via RG evolution, thus also contributing to the $0\nu\beta\beta$ decays. These loop contributions can be dominant in scenarios with hierarchical interactions to quarks, which provide further motivation for estimating their potential effects~\cite{Faroughy:2020ina}. Furthermore, as we will demonstrate, these loop-level constraints on dimension-seven operators can be far more stringent than those arising from tree-level meson decays.

The remainder of this paper is organized as follows. Our EFT framework is defined in Sec.~\ref{sec:eft-approach}, and the RG contributions that are relevant for $0\nu\beta\beta$ are systematically classified in Sec.~\ref{sec:flavor-mixing} through a logarithmic expansion. In Secs.~\ref{sec:pheno} and~\ref{sec:numerical}, we perform our phenomenological analysis by comparing tree-level constraints with those induced through RG evolution. Our findings are summarized in Sec.~\ref{sec:conclusion}.

\section{LNV operators at $d=7$}
\label{sec:eft-approach}

\begin{table*}[!t]
    \renewcommand{\arraystretch}{3.0}
    \centering
        \centering
        \begin{NiceTabular}{c|cccccc}
         $d=7$   & \;\;\;\;\;\; $\psi^2 H^4$ \;\;\;\;\;\; & \;\;\;\;\;\;$\psi^2 H^3 D$\;\;\;\;\;\;  &  \;\;\;\;\;\;$\psi^2 H^2 D^2$\;\;\;\;\;\;  &  \;\;\;\;\;\;$\psi^2 H^2 X$\;\;\;\;\;\; &  \;\;\;\;\;\;$\psi^4 H$\;\;\;\;\;\;  &  \;\;\;\;\;\;$\psi^4 D$\;\;\;\;\;\; \\\hline
       $\psi^2 H^4$ & $g^2,\,\lambda,\,y_q^2,\,{\color{gray}y_\ell^2}$ & $\color{gray}g^2 y_\ell,\,y_\ell^3$ & $g^4,\,\lambda g^2,\,\color{gray}\lambda y_\ell^2,\,y_\ell^4$ & $g^3\,,\color{gray}g y_\ell^2$ & \cellcolor{blue!5} ${\color{blue}\boldsymbol{\lambda y_q,\,y_q^3}},\,{\color{gray}\lambda y_\ell,\,y_\ell^3}$ & 0 \\ 
     $\psi^2 H^3 D$   & $0$ & $g^2,\,\lambda\,y_q^2,\,{\color{gray}y_\ell^2}$    & \color{gray}$g^2y_\ell,\,\lambda y_\ell,\,y_\ell^3$     &   0   &  
      \cellcolor{blue!5} $\color{blue}\boldsymbol{y_u y_d}$    &  $0$  \\ 
    $\psi^2 H^2 D^2$   & $0$ &  $0$    &   $g^2,\,\lambda,\,y_q^2,\,{\color{gray}y_\ell^2}$   &  $0$    &   $0$   & $ \cellcolor{blue!5} \color{blue}\boldsymbol{y_u y_d}$ \\ 
    $\psi^2 H^2 X$  & $0$ &   $0$   &   $g^3,\,{\color{gray}g y_\ell^2}$   &  $g^2,\,\lambda,\,y_q^2,\,{\color{gray}y_\ell^2}$    &  \cellcolor{blue!5} $\color{blue}\boldsymbol{g y_d}$   &   0  \\ 
    $\psi^4 H$   & $0$ & \cellcolor{blue!5} $\color{blue}\boldsymbol{y_u y_d}$   &  \cellcolor{blue!5} ${\color{blue}\boldsymbol{g^2 y_d}},\,{\color{gray}y_d y_\ell^2,\,y_u y_d y_\ell}$    &   \cellcolor{blue!5} ${\color{blue}\boldsymbol{g y_d}}$   & \cellcolor{blue!5} $g^2,\,\color{blue}\boldsymbol{y_q^2},\,{\color{gray}y_q y_\ell,\,y_\ell^2}$     & \cellcolor{blue!5} ${\color{blue}\boldsymbol{g^2 y_u}},\, {\color{gray}y_q y_\ell^2,\, y_q^2 y_\ell,\, y_\ell^3,\, g^2 y_\ell}$ \\ 
    $\psi^4 D$ &  $0$ &   $0$   &  \cellcolor{blue!5} $\color{blue}\boldsymbol{y_d y_u}$   &    $0$  &   $0$   & \cellcolor{blue!5}  ${\color{blue}\boldsymbol{y_q^2}}\,,g^2,\,{\color{gray}y_\ell^2}$\\
        \end{NiceTabular}
    \caption{\small \sl Form of the anomalous dimension matrix for dimension-seven SMEFT operators based on Ref.~\cite{Zhang:2023ndw}, where $g$ denotes the electroweak couplings $g_1$ or $g_2$, and the quark Yukawa coupling $y_q$ can be either $y_u$ or $y_d$. Dominant mixing effects for $0\nu\beta\beta$ decays are highlighted in blue, whereas those suppressed by lepton Yukawas are shown in gray.
    }
    \label{tab:SMEFT-Higgs-ope} 
\end{table*}

We assume that a \emph{mass gap} separates the electroweak scale from the scale of New Physics ($\Lambda$), so that the 
SMEFT provides the most convenient description of low-energy phenomena~\cite{Buchmuller:1985jz},
\begin{equation}
    \label{eq:smeft-d7}
    \mathcal{L}_{\rm EFT}= \mathcal{L}_{d \leq 4}+\sum_{d>4}\sum_a \dfrac{1}{\Lambda^{d-4}}\, \mathcal{C}_a^{(d)}\mathcal{O}_a^{(d)}\,,
\end{equation}

\noindent where $\mathcal{C}_a^{(d)}$ are the Wilson coefficients and $\mathcal{O}_a^{(d)}$ are the corresponding $d$-dimensional effective operators. The breaking of lepton number ($L$) by two numbers in the SMEFT arises through odd-dimension operators~\cite{Kobach:2016ami}, with the lowest-dimension operator being the Weinberg operator, defined in Eq.~\eqref{eq:Weinberg-operator}. The next-order operators that break $L$ appear at dimension-seven, with 12 independent operator structures that are collected in Table~\ref{tab:EFT-LNV-d7} in the operator basis of Ref.~\cite{Zhang:2023kvw}.~\footnote{Notice that Ref.~\cite{Zhang:2023kvw} and~\cite{Cirigliano:2017djv} adopt different operator bases. In particular, the dipole operators ($\psi^2 H^2 X$) are defined without the multiplication by the electroweak gauge couplings. Furthermore, the operator ${\mathcal{O}_{\bar{d}LueH}^{(7)}}$ is replaced by its Fierz equivalent, ${\mathcal{O}_{LeudH}^{(7)}=\epsilon_{ij}\,\big{(}L_i^T C\gamma_\mu e\big{)}\big{(}\bar{d}\gamma^\mu u\big{)}\, H_j}$, in Ref.~\cite{Cirigliano:2017djv}.} Our notation and conventions are spelled out in Appendix~\ref{app:notation}. In particular, we work in the down-aligned flavor basis for quarks, so that quark doublets read $Q_i = [(V^\dagger u_L)_i\,,~d_{Li}]^T$, where $i$ denotes the quark flavor index and $V$ stands for the Cabibbo–Kobayashi–Maskawa (CKM) matrix. Furthermore, we consider diagonal lepton Yukawas for convenience.

Although suppressed by an additional power of $v^2/\Lambda^2$ relative to $\smash{\mathcal{O}_{LH}^{(5)}}$, dimension-seven operators can be dominant in scenarios where they are generated at tree level, whereas the dimension-five Weinberg operator only arises at the loop level~\cite{Fridell:2023rtr}. In particular, the contributions from dimension-seven operators to $0\nu\beta\beta$ decays can be chirality-enhanced with respect to the standard neutrino-exchange contribution, as mentioned above~\cite{Cirigliano:2017djv,Cirigliano:2018yza}.

Previous analyses of dimension-seven operators in $0\nu\beta\beta$ decays have focused on the couplings to first-generation fermions, as they induce tree-level contributions to such processes. In these scenarios, current experimental limits from KamLAND-Zen~\cite{KamLAND-Zen:2024eml} allow us to test energy scales as large as $\mathcal{O}(300~\mathrm{TeV})$ in a perturbative regime \cite{Fajfer:2024uut}. Importantly, this level of sensitivity also allows us to probe suppressed contributions, such as those induced by operators coupled to heavy quarks, which would only appear at higher loop orders. These effects can be partially captured by the RG evolution of the EFT operators through the gauge and Yukawa interactions, as will be discussed in the following Section.

\section{Flavor mixing from RG Evolution}
\label{sec:flavor-mixing}

In this Section, we characterize the relevant operator-mixing effects for $0\nu\beta\beta$ decays and, in particular, the mixing of operators with different quark flavors. The RG equations describing the evolution of dimension-seven operators in the SMEFT can be written as follows,
\begin{equation}
    \label{eq:rge-d7}
    16\pi^2\,\mu\dfrac{\mathrm{d}}{\mathrm{d}\mu} \mathcal{C}^{(7)}_a = \gamma^{(7)}_{ba}\, \mathcal{C}_b^{(7)} \,+\, \dots \,,
\end{equation}
where the operator indices are denoted by $a,b$, and the ellipses encode the contributions from insertions of lower-dimensional operators~\cite{Zhang:2023kvw}, which are suppressed in the EFT scenarios that we will consider. The anomalous matrix $\gamma^{(7)}$ is fully known at one-loop order~\cite{Zhang:2023ndw}.~\footnote{See Ref.~\cite{Liao:2020vru} for a numerical implementation in a public library.}  

In this study, we are interested in the off-diagonal entries of $\gamma^{(7)}$, as they can generate RG-induced contributions to $0\nu\beta\beta$ decays from dimension-seven operators for which tree-level contributions are absent or suppressed; cf.~the highlighted operators in Table~\ref{tab:EFT-LNV-d7}. In other words, our RG analysis will allow us to constrain several operators that would otherwise be unconstrained at leading order. We stress that the relevant operator-mixing effects arise above the electroweak scale, as the only non-negligible running effects below the electroweak scale come from QCD evolution, which can change the magnitude of the scalar and tensor effective coefficients but will not mix different operators.

The solution to Eq.~\eqref{eq:rge-d7} for the running from $\Lambda$ to the electroweak scale $\mu_{\rm{ew}}\simeq m_W$ can be formally written as 
\begin{align}
\mathcal{C}_a^{(7)}(\mu_{\rm ew}) = U(\mu_{\rm ew},\Lambda)_{ab}\, \mathcal{C}_b^{(7)}(\Lambda)\,,
\end{align}

\noindent where $U(\mu_{\rm ew},\Lambda)$ is the evolution matrix, which takes the approximate form,
\begin{align}
    \label{eq:rge-d7-solution}
    U(\mu_{\rm ew},\Lambda)_{ab} &= \delta_{ab}+\dfrac{\gamma_{ba}^{(7)}}{16\pi^2} \log(\mu_{\rm ew}/\Lambda) \\*[0.35em]
    &+\dfrac{1}{2}\sum_c \dfrac{\gamma_{bc}^{(7)}\,\gamma_{ca}^{(7)}}{(16\pi^2)^2}\log^2(\mu_{\rm ew}/\Lambda)+\,\dots\,,\nonumber
\end{align}

\noindent where we have neglected the running of the SM couplings~\cite{Buras:2018gto}. By replacing $\gamma^{(7)}$ by its expression at one loop, we obtain the first leading-logarithm solution in the first term in the above equation, which is followed by the second leading-logarithmic contribution. The latter can be relevant if the direct one-loop mixing $(b)\to (a)$ vanishes, with a nonzero effect arising through the two-step mixing $(b)\to(c)\to (a)$.~\footnote{Examples of SMEFT operators for which the double-logarithm terms provide the leading contribution to a given process have been discussed, e.g., in the context of lepton flavor violation~\cite{EliasMiro:2021jgu,Plakias:2023esq}, electric dipole moments \cite{Cirigliano:2016njn, Kley:2021yhn}, and electroweak precision observables~\cite{Allwicher:2023shc,Allwicher:2023aql,Haisch:2024wnw,Born:2026tgm}.}

\begin{figure*}[!t]
    \centering
    \includegraphics[width=0.98\textwidth]{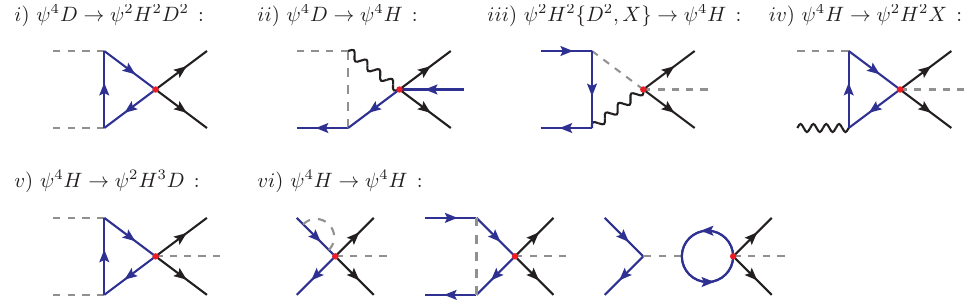}
    \caption{\small \sl Topologies of representative one-loop RG-induced contributions to $0\nu\beta\beta$ decays. Quark and lepton lines are denoted by blue and black lines, respectively. The insertion of dimension-seven SMEFT operators is depicted by the red dot. Notice, in particular, that the diagrams in panel $vi)$ can mix similar types of operators with different quark flavors.}
    \label{fig:rge-diagrams}
\end{figure*}

The contributions on the right-hand side of Eq.~\eqref{eq:rge-d7-solution} can arise from the mixing between different classes of operators, or from the mixing of the same type of operators but with different quark flavors, as illustrated in Fig.~\ref{fig:rge-diagrams}. The latter effects stem from the Yukawa-induced running, which introduces an irreducible source of quark flavor violation. Therefore, the running effects generated in this way are proportional to the CKM matrix elements. As it will be shown in Sec.~\ref{sec:pheno}, these are the most relevant RG effects for $0\nu\beta\beta$ decays.~\footnote{We have explicitly checked that the flavor misalignment induced by the running of the SM Yukawa matrices has a negligible effect on the scenarios that we study, see e.g.~Refs.~\cite{Coy:2019rfr,Aebischer:2020lsx,Aebischer:2025qhh}.}

The nonzero entries of the $\gamma^{(7)}$ matrix are summarized in Table~\ref{tab:SMEFT-Higgs-ope} in terms of the gauge couplings and SM Yukawas, with representative diagrams illustrated in Fig.~\ref{fig:rge-diagrams}. In this Table, the contributions that generate non-negligible mixing of operators that are relevant for $0\nu\beta\beta$ are highlighted in blue. These entries are proportional to the quark Yukawas, with the largest effects coming from operators with third-generation quarks. Contributions that are proportional to the lepton Yukawas are displayed in gray, as they are negligible for $\ell=e$, as we consider. Notably, all classes of operators mix into $\psi^4 H$ ones, which can give sizable contributions to $0\nu\beta\beta$ decays. 

In the following, we provide two examples to illustrate the operator-mixing effects described above, at different orders in the logarithmic expansion of Eq.~\eqref{eq:rge-d7-solution}.

\subsection*{Example I: Single-logarithm mixing}

Firstly, we consider the example of the operator $\mathcal{O}_{\bar{d}LueH}^{(7)}$ at the $\Lambda = \mathcal{O}(\mathrm{TeV})$ scale,
\begin{equation}
\mathcal{O}_{\substack{\bar{d}LueH\\ i \alpha  j \beta}}^{(7)} = \epsilon_{ab} \,\big{(}\bar{d}_i L^a_\alpha \big{)}\big{(}u^T_j C e_\beta\big{)} H^b\,,
\end{equation}

\noindent which is induced at tree level, for instance, in the scalar leptoquark scenario studied in Ref.~\cite{Fajfer:2024uut}. Lepton flavor indices will be fixed to $\alpha=\beta=1$. Furthermore, we assume that the coefficients related to first-generation quarks are suppressed by a given \emph{flavor ansatz}, so that tree-level contributions to $0\nu\beta\beta$ decays are absent. In this case, the dominant contributions to these processes arise from RG-induced mixing of the above operator into  $\mathcal{O}_{LeHD}$ ($\psi^2 H^3 D$) via the one-loop diagram illustrated in panel $\emph{v)}$ of Fig.~\ref{fig:rge-diagrams}~\cite{Zhang:2023ndw}, 
\begin{align}
\label{eq:rge-example-1}
16\pi^2 \dfrac{\mathrm{d}}{\mathrm{d}\log \mu
}\mathcal{C}_{\substack{LeHD\\ \alpha \alpha}}^{(7)} = -6 \, \mathcal{C}_{\substack{\bar{d}LueH\\ i\alpha j\alpha}}^{(7)}\,(y_u^\dagger\, y_d)_{ji} +\dots \,, 
\end{align}

\noindent where we have written only the non-diagonal terms that are not suppressed by lepton Yukawas, and we remind that
\begin{align}
\mathcal{O}_{\substack{LeHD\\ \alpha\beta}}^{(7)} &= \epsilon_{ab}\epsilon_{cd}\big{(}L^{aT}_\alpha C \gamma_\mu e_\beta \big{)}H^b H^c \big{(}i\,D^\mu H\big{)}^d\,.
\end{align}

\noindent Therefore, RG evolution will generate the operator $\mathcal{O}_{LeHD}^{(7)}$ at leading order, which modifies the $W$-boson couplings to leptons after electroweak-symmetry breaking, thus contributing to $0\nu\beta\beta$ decays. Technically, this operator induces $0\nu\beta\beta$ mechanisms that are suppressed by two powers in the chiral EFT power counting $\sim m_\pi^2/\Lambda_\chi^2$, where $\Lambda_\chi \sim 1$ GeV is the chiral-symmetry-breaking scale, but this is compensated by the large anomalous magnetic-moment of the nucleon~\cite{Cirigliano:2017djv} and the resulting $0\nu\beta\beta$ constraints are still severe. The CKM misalignment of the Yukawa matrices amounts to 
\begin{align}
\label{eq:rge-example-1}
y_u^\dagger y_d = \hat{y}_u^T\, V \, \hat{y}_d\,,
\end{align}

\noindent where $\hat{y}_d \equiv \mathrm{diag}(y_d,y_s,y_b)$ and $\hat{y}_u \equiv \mathrm{diag}(y_u,y_c,y_t)$, thus implying hierarchical contributions from Eq.~\eqref{eq:rge-example-1} to $0\nu\beta\beta$ decays (for $\alpha=\beta=1$), 
\begin{align}
    \mathcal{C}_{\substack{LeHD\\ \alpha\alpha}}^{(7)} &(\mu_{\mathrm{ew}})= \mathcal{C}_{\substack{LeHD\\ \alpha\alpha}}^{(7)}(\Lambda) \\[0.3em]
    &-\dfrac{3}{4\pi^2}\, \dfrac{m_{u_j} m_{d_i}}{v^2}V_{ji}\log \big{(}\mu_\mathrm{ew}/\Lambda\big{)}  \,\mathcal{C}_{\substack{\bar{d}LueH\\ i\alpha j\alpha}}^{(7)}(\Lambda) +\dots\,, \nonumber
\end{align}
where the ellipses represent additional RG-induced contributions. The largest effects come from third-generation quarks, which are proportional to $m_t\,m_b$, and for which the current limits from $0\nu\beta\beta$ decays can test scales up to $\mathcal{O}(30~\mathrm{TeV})$, as will be shown in our numerical analysis in Sec.~\ref{sec:numerical}.

\subsection*{Example II: Double-logarithm mixing}

The second example that we consider is the following $\psi^4D$ operator at the ultraviolet scale,
\begin{align}
\label{eq:ope-example-2}
\mathcal{O}_{\substack{\bar{d}uLLD\\ ij \alpha\beta}}^{(7)} = \epsilon_{ab} \big{(}\bar{d}_i \gamma_\mu u_j\big{)}\big{(}L^{aT}_\alpha C (iD^\mu L_\beta)^b\big{)} \,.
\end{align}

\noindent Similarly to the previous example, this operator contributes to $0\nu\beta\beta$ decays at tree level only for first-generation quark flavor indices, but coefficients involving heavy quark flavor can still affect these processes through RG evolution. As in the previous example, we will assume a flavor structure that suppresses the direct couplings to valence quarks.

The main difference from the previous example is that the leading RG contributions do not necessarily appear with the first leading logarithm, depending on the quark flavor indices. For instance, that is the case for $(i,j)=(3,1)$, for which the first logarithm to the $0\nu\beta\beta$ operators is suppressed by the up-quark Yukawa, thus making it smaller than the squared-log contributions. More specifically, one of the most relevant RG effects for the operator defined in Eq.~\eqref{eq:ope-example-2} is given by~\footnote{The one-loop mixing between $\mathcal{C}^{(7)}_{\bar{d}uLL D}$ and $\mathcal{C}^{(7)}_{\bar{d}LQLH1}$ also generates relevant contributions to $0\nu\beta\beta$ decays for specific quark flavor combinations, such as $(i,j)=(1,2)$ and $(i,j)=(1,3)$. These effects are not discussed above, for simplicity, but will be included in the numerical analysis in Sec.~\ref{sec:numerical}.}
\begin{align}
16\pi^2 \dfrac{\mathrm{d}}{\mathrm{d}\,\log \mu} \mathcal{C}_{\substack{LHD1\\ \alpha\alpha}}^{(7)} &=  12\, (y_u^\dagger y_d)_{ji}\,\mathcal{C}_{\substack{\bar{d}u LLD\\ ij \alpha\alpha}}^{(7)}+\dots\,, 
\end{align}

\noindent which arises from diagram \emph{i)} in Fig.~\ref{fig:rge-diagrams}, and gives the leading contribution to $0\nu\beta\beta$ decays, e.g., for $(i,j)=(3,3)$. However, this contribution is suppressed for operators with first and second-generation quarks, since $(y_u^\dagger y_d)_{ji}= y_{u_j}\,y_{d_i} V_{ji}$.

For operators with suppressed first leading-logarithm contributions, the double logarithms can be relevant. In particular, 
\begin{align}
16\pi^2 &\dfrac{\mathrm{d}}{\mathrm{d}\,\log \mu} \mathcal{C}_{\substack{\bar{Q}uLLH\\ ij \alpha\alpha}}^{(7)} = 3 g_2^2 (y_d)_{ik} \, \,\mathcal{C}_{\substack{\bar{d}u LLD\\ kj \alpha\alpha}}^{(7)}  \\*
&+\dfrac{5}{2}(y_u y_u^\dagger)_{ik}\,\mathcal{C}_{\substack{\bar{Q}uLLH\\ kj \alpha\alpha }}^{(7)} + 6 (y_u^\dagger)_{lk} (y_u)_{ij} \,\mathcal{C}_{\substack{\bar{Q}uLLH \\ kl\alpha\alpha}}^{(7)} + \dots\,,\nonumber
\end{align}

\noindent where we have retained only the largest non-diagonal terms for the scenarios we are interested in. These effects are induced by diagram \emph{ii)} in Fig.~\ref{fig:rge-diagrams}, which thus contribute to $0\nu\beta\beta$ decays via the two-step mixing, 
\begin{align}
\mathcal{C}_{\substack{\bar{d}uLLD\\i j\alpha\alpha}}\; \to \;\mathcal{C}_{\substack{\bar{Q}uLLH\\3j\alpha\alpha}} \; \to \;\;\mathcal{C}_{\substack{\bar{Q} u LLH\\ij\alpha\alpha}}\,,
\end{align}
where the second step corresponds to the flavor mixing of the same type of operator into the one with light-quark flavors. More explicitly, for $(i,j)=(3,1)$,
\begin{align}
    &\mathcal{C}^{(7)}_{\substack{\bar{Q}uLLH\\ 11\alpha\alpha}}(\mu_{\mathrm{ew}}) = \mathcal{C}^{(7)}_{\substack{\bar{Q}uLLH\\ 11\alpha\alpha}}(\Lambda)\\*[0.3em]
    &\qquad+\dfrac{15}{4} \dfrac{g_2^2\, y_b\, y_t^2}{(16\pi^2)^2} V_{tb } V_{td}^\ast \log^2 \big{(} \mu_{\mathrm{ew}}/\Lambda\big{)} \, \mathcal{C}^{(7)}_{\substack{\bar{d}uLLD\\31 \alpha\alpha}} (\Lambda)+\dots\,, \nonumber
\end{align}

\noindent where we have kept only the bottom- and top-quark Yukawas, for simplicity. These RG contributions lead to the dominant effects to $0\nu\beta\beta$ decays at low energies for this specific flavor combination, allowing us to test scales as large as $\mathcal{O}(1~\mathrm{TeV})$ with current data, as will be shown in Sec.~\ref{sec:pheno}. Similar conclusions hold for the $(i,j)=(2,1)$ operator. 

The above example highlights the importance of considering contributions beyond the first leading-logarithm in phenomenological analyses. In Sec.~\ref{sec:numerical}, we will resum these contributions, while keeping track of the order in the expansion of Eq.~\eqref{eq:rge-d7-solution} at which the leading effects to $0\nu\beta\beta$ decays arise.

\section{Phenomenological constraints}
\label{sec:pheno}

\begin{figure*}[!t]
    \includegraphics[width=1.\textwidth]{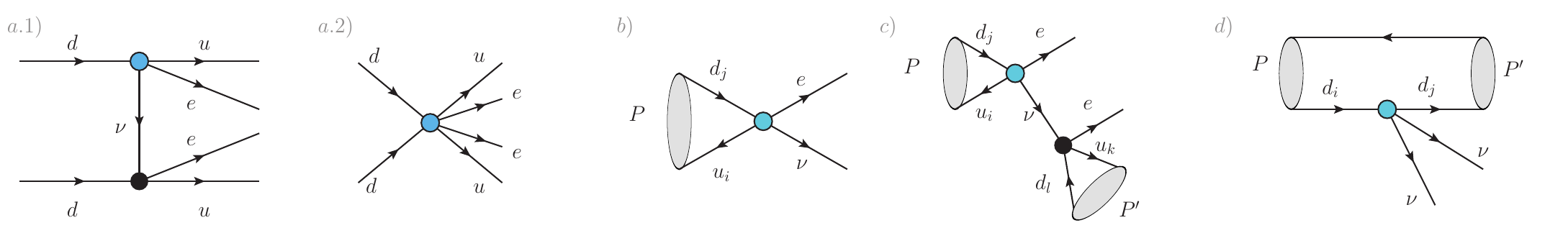}
    \caption{\small \sl Phenomenological constraints on LNV interactions from $0\nu\beta\beta$ decays, and from the different types of meson decays, namely $P^-\to e +\mathrm{inv}$, $P^-\to P^{\prime +} e^- e^-$  and $P\to P^\prime +\mathrm{inv}$, where $P^{(\prime)}$ denote pseudoscalar mesons. The blue circles represent insertions of LNV low-energy operators, whereas the insertion of the SM weak vertex is depicted by the black circle. Notice, in particular, that the LNV operators entering $0\nu\beta\beta$ in diagrams $a.1)$  and $a.2)$ can be loop-induced in our setup.}
    \label{fig:diagrams-flavor}
\end{figure*}

In this Section, we estimate the constraints from $0\nu\beta\beta$ decays on the full set of SMEFT operators collected in Table~\ref{tab:EFT-LNV-d7} by including one-loop RG effects. In particular, we consider operators with all possible quark flavors, comparing these constraints with those obtained from
tree-level meson decays in the kaon, and $D$- and $B$-meson sectors, as illustrated in Fig.~\ref{fig:diagrams-flavor}.

\subsection{Low-energy EFT for $0\nu\beta\beta$}

Firstly, we introduce the Low-Energy EFT (LEFT) Lagrangian needed for describing processes that violate Lepton Number by two units ($\Delta L=2$) at low-energy scales,
\begin{align}
\label{eq:left-lnv}
\mathcal{L}_{\Delta L=2}^{\mathrm{LEFT}} = \mathcal{L}_{\Delta L=2}^{(3)}+\mathcal{L}_{\Delta L=2}^{(6)}+\mathcal{L}_{\Delta L=2}^{(7)}+\dots\,,
\end{align}

\noindent where the operators are invariant under $SU(3)_c\times U(1)_{\mathrm{em}}$. The first term in Eq.~\eqref{eq:left-lnv} corresponds to Majorana neutrino masses, 
\begin{align}
     \mathcal{L}_{\Delta L=2}^{(3)} = - \dfrac{1}{2} \big{(}m_\nu\big{)}_{\alpha \beta} \, {\nu}_{L \alpha}^T C \nu_{L \beta}\,,
\end{align}
where the neutrino mass matrix $m_\nu = \mathcal{O}(v^2/\Lambda)$ is induced by the Weinberg operator defined in Eq.~\eqref{eq:Weinberg-operator}, and we remind that $C=i\gamma_2\gamma_0$ denotes the charge conjugation matrix. The effective Lagrangian with dimension $d \geq 6$ operators can be expressed as 
\begin{align}
    \label{eq:left}
    \mathcal{L}_{\Delta L=2}^{(d)} =  \dfrac{1}{v^{d-4}}
 \sum_a C_a^{(d)}\, O_a^{(d)}+\mathrm{h.c.}\,,
\end{align}
where $v = (\sqrt{2}G_F)^{-1/2}\approx 246$~GeV is the electroweak vacuum expectation value, $G_F$ denotes the Fermi constant, and $\smash{C_a^{(d)}}$ stands for the Wilson coefficients of the $d$-dimensional operators $\smash{O_a^{(d)}}$, which are defined below. In the following, we fix the lepton and neutrino flavors to the first generation in our LEFT basis, but leave generic flavor indices for quarks, unless stated otherwise.

\paragraph*{$d=6$\,:} The most relevant low-energy operators in our analysis are the dimension-six semileptonic ones, 
\begin{align}
    \label{eq:left-d6}
O_{\substack{V_L \\  ij}}^{(6)} &= \big{(}\bar{u}_{Li} \gamma^\mu d_{Lj}\big{)}\big{(} \bar{e}_{R} \gamma_\mu C\bar{\nu}_{Le}^T\big{)}\,, \\*[0.3em]
O_{\substack{V_R\\ ij}}^{(6)} &= \big{(}\bar{u}_{Ri} \gamma^\mu d_{Rj}\big{)}\big{(} \bar{e}_{R} \gamma_\mu C\bar{\nu}_{Le}^T\big{)}\,, \nonumber\\*[0.3em]
O_{\substack{S_L\\ ij}}^{(6)} &= \big{(}\bar{u}_{Ri}  d_{Lj}\big{)}\big{(} \bar{e}_{L}  C\bar{\nu}_{Le}^T\big{)}\,,\nonumber\\*[0.3em]
O_{\substack{S_R\\ ij}}^{(6)} &= \big{(}\bar{u}_{Li}  d_{Rj}\big{)}\big{(} \bar{e}_{L }  C\bar{\nu}_{Le}^T\big{)}\,,\nonumber\\*[0.3em]
O_{\substack{T\\ ij}}^{(6)} &= \big{(}\bar{u}_{Li} \sigma^{\mu\nu} d_{Rj}\big{)}\big{(} \bar{e}_{L} \sigma^{\mu\nu}  C\bar{\nu}_{Le}^T\big{)}\,,\nonumber
\end{align}

\noindent where $i,j$ denote quark flavor indices, as before. These coefficients are induced by the dimension-seven SMEFT operators given in Eq.~\eqref{eq:smeft-d7}, so that $C^{(6)}=\mathcal{O}(v^3/\Lambda^3)$. The complete expressions for the tree-level matching for these coefficients are given in Appendix~\ref{app:matching}.

\paragraph*{$d=7$\,:} There are specific SMEFT operators that contribute to $C^{(7)}$, but not to $C^{(6)}$. For this reason, it is essential to extend the LEFT basis. An example is the operator $\smash{\mathcal{O}_{L H D 1}^{(7)}}$, which generates a four-fermion operator with a derivative after integrating out the SM gauge bosons, cf.~Eqs.~\eqref{eq:d6-matching-1}-\eqref{eq:d6-matching-2} in Appendix~\ref{app:matching}. For such operators, the leading effects at low-energy scales are captured by the following dimension-seven operators~\cite{Cirigliano:2017djv},
\begin{align}
    \label{eq:left-d7}
    O_{\substack{V_L}}^{(7)} &= \big{(}\bar{u}_{L}\gamma^\mu d_{L}\big{)}\big{(}\bar{e}_{L}\overleftrightarrow{\partial}_\mu C\bar{\nu}_{Le}^T\big{)}\,,\\*[0.35em]
    O_{\substack{V_R}}^{(7)} &=\big{(}\bar{u}_{R}\gamma^\mu d_{R}\big{)}\big{(}\bar{e}_{L}\overleftrightarrow{\partial}_\mu C\bar{\nu}_{Le}^T\big{)}\,,\nonumber
\end{align}

\noindent where $C^{(7)}=\mathcal{O}(v^3/\Lambda^3)$ and we have only kept the operators that are induced at tree level by $d=7$ operators in the SMEFT~\cite{Cirigliano:2017djv}. We restrict ourselves to the first generation of quarks in Eq.~\eqref{eq:left-d7}, as these are the most relevant $d=7$ LEFT operators under our assumptions.

\paragraph*{$d=9$\,:}The same operators in the SMEFT that generate Eq.~\eqref{eq:left-d7} also contribute to $d=9$ low-energy operators, after integrating out two gauge boson propagators~\cite{Cirigliano:2017djv},
\begin{align}
    \label{eq:left-d9}
    O_{\substack{1 }}^{(9)} &= \big{(} \bar{u}_L \gamma^\mu d_L \big{)}  \big{(} \bar{u}_L \gamma^\mu d_L \big{)} \big{(} \bar{e}_{L} C e_{L}\big{)}\,, \\[0.4em]
    O_{\substack{4}}^{(9)} &= \big{(} \bar{u}_L \gamma^\mu d_L \big{)}  \big{(} \bar{u}_R \gamma^\mu d_R \big{)} \big{(} \bar{e}_{L} C e_{L}\big{)}\,, \nonumber
\end{align}
where $C^{(9)}=\mathcal{O}(v^3/\Lambda^3)$, thus with the same suppression on the ultraviolet scale. Once again, we only consider the first generation of quarks for these operators, which are numerically relevant for the analysis of $0\nu\beta\beta$ decays. 

\

Finally, we remind that to derive constraints on the effective coefficients, the QCD running below and above the electroweak scale must be taken into account, which impacts scalar and tensor operators and is known up to four-loops~\cite{Vermaseren:1997fq,Chetyrkin:1997dh,Gracey:2022vqr,vanRitbergen:1997va,Chetyrkin:1997sg}. For consistency with out SMEFT analysis, we will consider the one-loop QCD evolution, since higher-order corrections have only a small impact on our results.

\subsection{$0\nu\beta\beta$ decays}

The computation of $0\nu\beta\beta$ half-lives has been a longstanding problem. In the last decade, progress has been made in describing nuclear LNV using EFT methods~\cite{Cirigliano:2022oqy}. Slightly above the scale where QCD becomes non-perturbative, LNV can be described by the LEFT operators appearing in Eq. \eqref{eq:left-lnv}. At a lower energy scale, the LEFT operators can be matched to LNV operators in chiral perturbation theory. In turn, these chiral operators can be used to derive LNV nuclear $0\nu\beta\beta$ transition operators, often called neutrino potentials. The transition operators must be inserted into nuclear wave functions describing the initial- and final-state atomic nuclei, a step that required nuclear many-body theory. Finally, the half-lives depend on atomic phase space factors that depend on the reaction $Q$ value and the leptonic structure of the decay. 

The above steps are efficiently described by a cascade of EFTs, each valid in its own energy range, leading to a `Master Formula'  \cite{Cirigliano:2017djv,Cirigliano:2018yza} which has been automized in the Python tool $\nu$DoBe \cite{Scholer:2023bnn}. $
\nu$DoBe can be used to directly compute $0\nu\beta\beta$ rates as a function of SMEFT or LEFT Wilson coefficients. 

In our numerical analysis, we consider the experimental limit on the $0\nu\beta\beta$ half-life $\smash{T_{1/2}^{0\nu}>3.8
\times 10^{26}~\mathrm{yr}}$ ($90\%$~CL), which was obtained by KamLAND-Zen for $^{136}\mathrm{Xe}$~\cite{KamLAND-Zen:2024eml}. The resulting constraints on the $d=7$ operators are collected in Sec.~\ref{sec:numerical}, where we will make the comparison to meson decays. Furthermore, we will explore the interplay of tree- and loop-level constraints, relying on the logarithmic expansion introduced in Eq.~\eqref{eq:rge-d7-solution}.

\subsection{Meson decays}

The LEFT operators introduced in Eq.~\eqref{eq:left-d6} can affect charged- and neutral-current meson decays, depending on the specific operator. In the following, we derive constraints from these processes, which will be compared to the RG bounds from $0\nu\beta\beta$ decays in Sec.~\ref{sec:numerical}.

\paragraph*{$P\to \ell \nu$} The simplest processes for constraining operators of the type $\bar{u}d \,\bar{\ell} \nu^C$ are charged-current leptonic decays of pseudoscalar mesons, $P\to \ell {\nu}$, as depicted in panel \emph{b)}   of Fig.~\ref{fig:diagrams-flavor}. These are not strict LNV processes, but neutrinos are not observed experimentally, so that the signal receives contributions from operators with both $\nu_L$ and $\nu_L^C$. The latter induces contributions that do not interfere with the SM ones,
\begin{align}
    \mathcal{B}(P\to e+ \mathrm{inv}) \equiv \mathcal{B}(P\to e \nu_\ell)^{\mathrm{SM}}\,\Big{[}1+ |\delta_{Pe_2}|^2\Big{]}\,,
\end{align}

\noindent where $\delta_{Pe_2}$ is defined for the $u_i\to d_j e \nu$ transition as follows, 
\begin{align}
    \delta_{Pe_2} = \dfrac{1}{2|V_{ij}|}&\bigg{[}C_{\substack{V_R\\ij}}^{(6)}-C_{\substack{S_R\\ij}}^{(6)}\, \dfrac{m_P^2}{m_e(m_{u_i}+m_{d_j})}    \bigg{]}- (L\leftrightarrow R) \,,
\end{align}

\noindent where $m_P$ is the $P$-meson mass, $m_{u_i}$ and $m_{d_j}$ denote the relevant quark masses, and neutrino masses have been neglected. We have only kept the contributions from LNV operators.~\footnote{See, e.g., Ref.~\cite{Becirevic:2020rzi} for complete expressions.} The relevant processes for the comparison to $0\nu\beta\beta$ are decays with electrons in the final state, namely $\pi\to e\nu$, $K\to e\nu$, $\smash{D_{(s)}\to e\nu}$ and $B\to e\nu$~\cite{ParticleDataGroup:2024cfk}. Semileptonic decays could also be used to constrain such operators, with the advantage of being sensitive to operators with different Lorentz structures~\cite{Becirevic:2020rzi}. However, they would lead to relatively weak bounds, which we opt not to include in our analysis.

In our numerical analysis, we consider the experimental averages provided in PDG~\cite{ParticleDataGroup:2024cfk} and the decay constants collected in FLAG~\cite{FlavourLatticeAveragingGroupFLAG:2024oxs}.  By using these inputs, we can test the relevant dimension-seven coefficients with an effective scale $\Lambda\simeq 9 \,{\rm TeV}$ in the kaon sector and $\Lambda\simeq 4 \,{\rm TeV}$ with $D$- and $B$-meson decays, for coefficients ${C}^{(6)}\simeq v^3/\Lambda^3$, i.e., for order unity couplings in the ultraviolet. These constraints will be compared to the other bounds for all dimension-seven operators in Sec.~\ref{sec:numerical}.

\paragraph*{$P^-\to P^{\prime\,+} \ell^- \ell^-$} We briefly comment on three-body LNV decays, which can be induced by the insertion of the operators in Eq.~\eqref{eq:left-d6} and the SM weak-interaction vertex, as illustrated in panel \emph{c)} of Fig.~\ref{fig:diagrams-flavor}. The decay rates for these processes can be schematically written as follows,
\begin{equation}
    \Gamma(P^-\to P^{\prime\,+} \ell^- \ell^-) 
    \simeq \dfrac{f_P^2 f_{P^\prime}^2 }{(16\pi)^3}\, m_P^5 G_F^4 |\mathcal{V}|^2\,|{C}^{(6)}|^2\,,
\end{equation}

\noindent where we have used the vacuum insertion approximation, and the masses of final-state particles are neglected, for simplicity. In this expression, $f_{P^{(\prime)}}$ denotes the decay constants of the $P^{(\prime)}$ mesons, and $\mathcal{V}$ stands for the relevant CKM matrix element arising from the insertion of the weak vertex. The complete decay rate expressions for this process are provided in Appendix~\ref{app:lnv-meson-formulas}, which will be used in our numerical analysis.

By using the experimental limits provided in PDG~\cite{ParticleDataGroup:2024cfk}, we are able to test scales $\Lambda \lesssim {250}$ GeV for effective coefficients ${C}^{(6)}\simeq v^3/\Lambda^3$. The weaker sensitivity compared to the processes discussed above stems from the double EFT insertion and the three-body phase space, which highly suppresses the decay rates. Notice, however, that such processes are very efficient probes of heavy neutral leptons with masses $m_N$ in the range $m_N \in (2 m_{\ell}\,, m_P-m_{P^\prime})$, due to the resonant contributions $P^- \to \ell^- N (\to P^{\prime+} \ell^-)$ that are possible in these scenarios, see Ref.~\cite{Abada:2017jjx} and references therein.

\paragraph*{$P\to P^{\prime} \nu \nu$} Finally, we discuss the flavor-changing neutral-current processes $P\to P^{\prime}\nu\nu$, where $P^{(\prime)}$ are pseudoscalar mesons, as illustrated in panel \emph{d)} of Fig.~\ref{fig:diagrams-flavor}. These channels can specifically probe the $\smash{\mathcal{O}_{\bar{d}LQLH1}^{(7)}}$ and $\smash{\mathcal{O}_{\bar{Q}uLLH}^{(7)}}$ operators at tree level, as they also contribute to down- and up-type neutral-current transitions after electroweak symmetry breaking, respectively. The relevant LEFT Lagrangian reads~\cite{Fridell:2023rtr,Buras:2024ewl},
\begin{align}
    \mathcal{L}_{\Delta L=2}^{(6)} \supset \dfrac{1}{v^2} &\bigg{[} D_{\substack{T_L\\ ij \alpha\beta}}^{(6)} \big{(}\bar{q}_{Ri} \sigma^{\mu\nu} q_{Lj}\big{)}\big{(}\overline{\nu^C_{L\alpha}}\sigma^{\mu\nu} \nu_{L\beta} \big{)}\\*[0.35em]
    &+D_{\substack{S_L\\ ij \alpha\beta}}^{(6)} \big{(}\bar{q}_{Ri} q_{Lj}\big{)}\big{(}\overline{\nu^C_{L\alpha}}\nu_{L\beta} \big{)}\nonumber \\*[0.35em]
    &+D_{\substack{S_R\\ ij \alpha\beta}}^{(6)} \big{(}\bar{q}_{Li} q_{Rj}\big{)}\big{(}\overline{\nu^C_{L\alpha}}\nu_{L\beta} \big{)}\bigg{]    }  + \mathrm{h.c.} \,,   \nonumber
\end{align}
where $q=u,d$, and the low-energy coefficients are denoted by $D_a^{(6)}$ to distinguish them from the ones entering charged-currents. For the down-type transition, these coefficients are given at the electroweak scale by
\begin{align}
    D_{\substack{S_L\\ ij \alpha\beta}}^{(6)} &\overset{(q=d)}{=}  -\dfrac{1}{2\sqrt{2}}\dfrac{v^3}{\Lambda^3} \, \mathcal{C}_{\substack{{\bar{d}LQLH1}\\ i\beta j\alpha}}^{(7)} + (\alpha\leftrightarrow \beta) \,,\\[0.35em]
    \, D_{\substack{T_L\\ ij \alpha\beta}}^{(6)} &\overset{(q=d)}{=} -\dfrac{1}{8\sqrt{2}}\dfrac{v^3}{\Lambda^3} \, \mathcal{C}_{\substack{{\bar{d}LQLH1}\\ i\beta j\alpha}}^{(7)} - (\alpha\leftrightarrow \beta) \,, \nonumber
\end{align}
whereas only the following scalar coefficient is nonzero for up-type transitions,
\begin{align}
    D_{\substack{S_R\\ ij \alpha\beta}}^{(6)} &\overset{(q=u)}{=}  \dfrac{v^3}{\Lambda^3}\, V_{ik} \, \mathcal{C}_{\substack{\bar{Q}uLLH\\ kj\alpha\beta}}^{(7)} +(\alpha\leftrightarrow \beta)\,.
\end{align}
Focusing on diagonal $\alpha=\beta$ flavors, only the scalar coefficient is nonzero, with the following contribution to the $P\to P^\prime \nu_\alpha \nu_\alpha$ decay rate~\cite{Felkl:2021uxi,Felkl:2023ayn},
\begin{align}
    \dfrac{\mathrm{d}\Gamma}{\mathrm{d}q^2}(P &\to P^\prime + \mathrm{inv})_{\mathrm{NP}} = \dfrac{G_F^2 q^2 \lambda_{P}^{1/2}}{256 \pi^3 m_P^3}\\*
    &\qquad \times \bigg{(}\dfrac{m_P^2-m_{P^\prime}^2}{m_{q_i}-m_{q_j}}\bigg{)}^2 \big{[}f_0(q^2)\big{]}^2\,\Big{|}D_{\substack{S_L\\ ij \alpha\alpha}}^{(6)}+D_{\substack{S_R\\ ij \alpha\alpha}}^{(6)}\Big{|}^2 \,, \nonumber
\end{align}

\noindent where we consider the $q_i\to q_j \nu \nu$ transition, $m_{P^{(\prime)}}$ denotes the $P^{(\prime)}$ meson mass, $q^2$ is the di-neutrino invariant-mass squared, $\lambda_P\equiv \lambda(q^2,m_P^2,m_{P^\prime}^2)$, with $\lambda(a^2,b^2,c^2) = (a^2-(b-c)^2)(a^2-(b+c)^2)$, and $f_0$ stands for the $P\to P^\prime$ scalar form-factor~\cite{FlavourLatticeAveragingGroupFLAG:2024oxs}. Notice that the contribution given above does not interfere with the SM one. Furthermore, we stress again that we account in our numerical analysis for the QCD evolution below and above the electroweak scale~\cite{Vermaseren:1997fq,Chetyrkin:1997dh,Gracey:2022vqr,vanRitbergen:1997va,Chetyrkin:1997sg}.

The most relevant experimental channels are $K\to \pi + \mathrm{inv}$ and $B\to K+\mathrm{inv}$, which have been recently observed by NA62~\cite{NA62:2024pjp} and Belle-II~\cite{Belle-II:2023esi}, respectively. Experimental limits are also available for the $D\to\pi+\mathrm{inv}$~\cite{BESIII:2021slf} and $B\to \pi +\mathrm{inv}$~\cite{Belle:2017oht} decays, which can test different quark flavor transitions~\cite{Bause:2021cna,Hiller:2025zgr}. We find that the best limits are given by kaon observables, which can test scales up to $\Lambda \simeq 40~\mathrm{TeV}$. In the $B$-sector, Belle-II observed a mild excess of  $B^+ \to K^+ + \rm inv$ with respect to the SM predictions for $B^+ \to K^+ \nu \bar{\nu}$, which can be translated into an effective scale of $\Lambda \simeq 3~\mathrm{TeV}$ for $\mathcal{O}(1)$ couplings~\cite{Abada:2026dlb}, cf.~Fig.~\ref{fig:eft-limits-quark}.

\begin{figure}[!t]
    \includegraphics[width=0.48\textwidth]{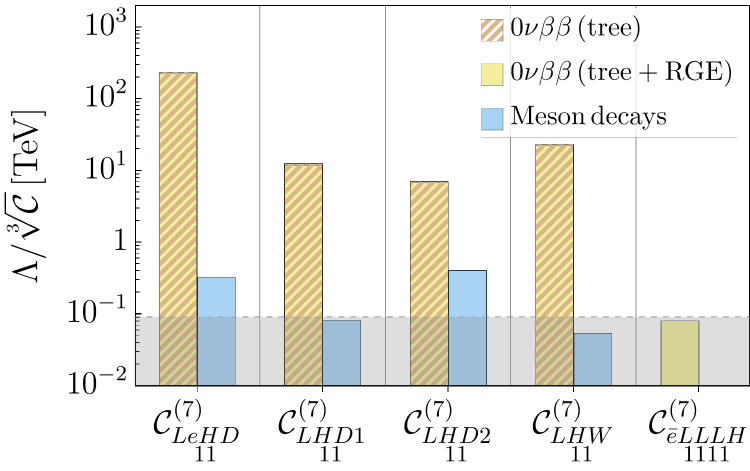}
    \caption{\small \sl Limits on $\Lambda/\sqrt[3]{\mathcal{C}}$ for dimension-seven LNV operators without quark fields obtained from $0\nu\beta\beta$ decays at tree (orange) and loop level (yellow), and from meson decays (blue). The gray hatched area corresponds to the region where $\Lambda < m_Z$ for perturbative coefficients ($\mathcal{C} \lesssim {4\pi}$), in which the EFT approach cannot be consistently used.}
    \label{fig:eft-limits}
\end{figure}

\section{Numerical analysis}
\label{sec:numerical}

\begin{figure*}[!p]
    \includegraphics[width=1\textwidth]{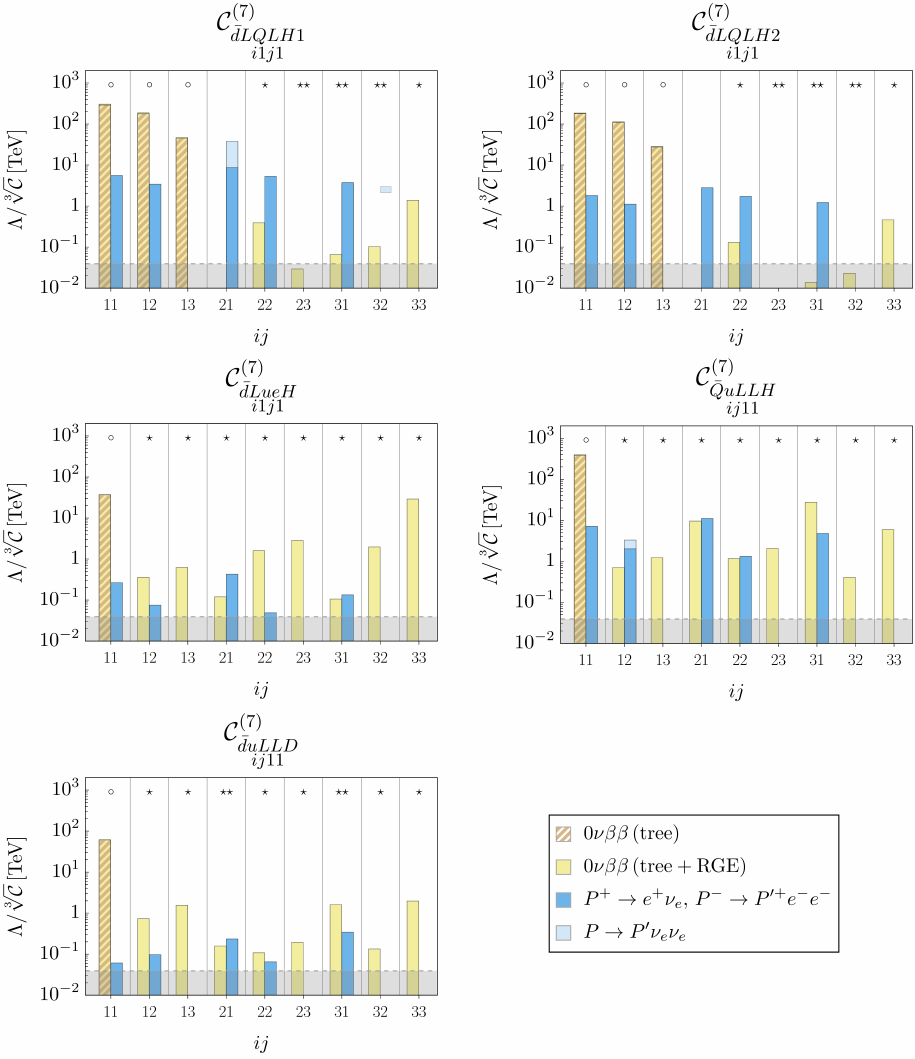}
    \caption{\small \sl  Constraints on $\Lambda/\sqrt[3]{\mathcal{C}}$ for dimension-seven LNV operators of type $\psi^4 H$ and $\psi^4 D$ containing quark fields obtained from $0\nu\beta\beta$ decays at tree (orange) and loop level (yellow), and from meson decays (blue). Quark flavor indices are denoted by $i,j \in \lbrace 1,2,3 \rbrace$. The upper symbols represent the order in the logarithmic expansion from Eq.~(\ref{eq:rge-d7-solution}) at which the leading effect appears, namely, at tree level ($\circ$), first leading-logarithm ($\star$), and second leading-logarithm ($\star\star$). The gray hatched area corresponds to the region where $\Lambda < m_Z$ for perturbative coefficients ($\mathcal{C} \lesssim {4\pi}$), in which the EFT approach cannot be consistently used.}
    \label{fig:eft-limits-quark}
\end{figure*}

In this Section, we compare the various probes of dimension-seven operators discussed above, as depicted in Fig.~\ref{fig:diagrams-flavor}. In particular, we study in detail the impact of the RG effects discussed in Sec.~\ref{sec:flavor-mixing} on the constraints derived from $0\nu\beta\beta$ decays.~\footnote{The RG-evolution effects that are independent of flavor mixing have been previously discussed in Ref.~\cite{Graf:2025cfk}, including the effects of one-loop matching to concrete models. These effects have also been automatized in Ref.~\cite{Liao:2025lxg}.} Following Sec.~\ref{sec:pheno}, we fix the lepton indices to the first generation, as these are the only flavors accessible with  $0\nu\beta\beta$ decays.

In Fig.~\ref{fig:eft-limits}, we present the limits on the operators without quark fields, namely those of types $\psi^2 H^3 D$, $\psi^2 H^2 D^2$ and $\psi^2 H^2 X$, as well as the purely leptonic $\psi^4 H$ operator. For most of these operators, we find that tree-level contributions to $0\nu\beta\beta$ decays are dominant, leading to limits at least one order of magnitude more stringent than those obtained from meson decays. These tree-level effects arise through the modifications of the $W$-boson interactions with leptons. The only exception is the leptonic operator $\mathcal{C}_{\bar{e}LLL H}$, for which RG-induced contributions are dominant, but suppressed by the lepton Yukawas. Therefore, given the current experimental precision, these effects that are far too small to be consistently probed with the EFT approach, as indicated by the shaded gray area in Fig.~\ref{fig:eft-limits}.

In Fig.~\ref{fig:eft-limits-quark}, we study the operators of $\psi^4 H$ and $\psi^4D$ containing quark fields, for which RG effects can be dominant depending on the quark flavors. That is the case for operators with heavy quarks, which do not contribute at tree level to $0\nu\beta\beta$ decays.
In particular, the relevant RG effects are proportional to the quark Yukawas, being larger for the third generation.
From Fig.~\ref{fig:eft-limits-quark}, we find that the combination of tree- and loop-level effects in $0\nu\beta\beta$ decays is, in most cases, more constraining than the bounds derived from the different types of meson decays.

To derive the RG effects mentioned above, we have also considered contributions beyond the first leading-logarithm in~Eq.~\eqref{eq:rge-d7-solution}. For most operators, the leading RG effects come from the first leading-logarithm, as indicated by the $(\star)$ symbol in Fig.~\ref{fig:eft-limits-quark}. However, there are operators for which the second leading-logarithm dominates and thus provides the most stringent constraints, as depicted by the $(\star\star)$ symbol. For instance, this is the case for $\smash{\mathcal{C}_{\bar{d}uLLD}^{(7)}}$ with flavor indices $(i,j)=(2,1)$ and $(3,1)$, as single logarithm mixing is suppressed for this operator, as discussed in Sec.~\ref{sec:flavor-mixing}. Other examples are $\smash{\mathcal{C}_{\bar{d}LQLH1}^{(7)}}$ and $\mathcal{C}_{\bar{d}LQLH2}^{(7)}$ also for specific flavor combinations, such as $(i,j)=(3,1)$ and $(3,2)$, as depicted in Fig.~\ref{fig:eft-limits-quark}.

Interestingly, the operator $\mathcal{O}_{\bar{d}LueH}^{(7)}$ represents a case in which the constraint obtained through loop effects for flavor indices $(i,j)=(3,3)$ is comparable to those obtained at tree level for valence quarks. As discussed in Sec.~\ref{sec:flavor-mixing}, for third-family quark indices this operator can mix into $\smash{\mathcal{O}_{LeHD}^{(7)}}$, which modifies the $W$-boson coupling to left-handed leptons. While this RG-induced contribution is suppressed by a loop factor, its contribution to $0\nu\beta\beta$ decay benefits from a milder suppression in the chiral expansion when compared to the one obtained when $(i,j)=(1,1)$, which generates right-handed couplings. This compensation allows the RG-induced constraints to become comparable to the tree-level limits.

Finally, we briefly comment on the comparison between our results with those from Ref.~\cite{Liao:2025lxg}, in which the running of LNV has been automated in a numerical tool. We find an overall good agreement for most operators, except for those that contribute to the Weinberg operators $\smash{\mathcal{O}_{LH}^{(5)}}$ and $\smash{\mathcal{O}_{LH}^{(7)}}$ though RGEs, for which their constraints are orders of magnitude more stringent than ours. We traced back this disagreement to unrealistic neutrino masses in their calculation, which modify the Weinberg operator contributions to $0\nu\beta\beta$ decays.~\footnote{An example is the operator $\mathcal{C}_{dlqlH1}^{(7)}$, with quark flavor indices $i=j=3$, which generates a contribution to the Weinberg operator proportional to $y_b$. } Indeed, several operators induce sizable contributions to neutrino masses~\cite{Chala:2021juk}, which must be canceled out by other EFT contributions, to ensure that active neutrino masses remain compatible with current experimental bounds, i.e.~$m_\nu \lesssim 0.1~\mathrm{eV}$.

\section{Conclusion}
\label{sec:conclusion}

In this paper, we have performed a complete Renormalization Group  analysis of $0\nu\beta\beta$ decays within the Standard Model Effective Field Theory, including operators up to dimension seven. Such operators can arise in radiative models of neutrino masses and can lead to sizable contributions to $0\nu\beta\beta$ decays when first-generation fermions are involved. By relying on the one-loop anomalous-dimension matrix computed in Ref.~\cite{Zhang:2023ndw}, we have shown that many operators that do not contribute to these processes at tree level can nevertheless induce significant effects through RG evolution. This is particularly relevant for operators involving heavy quark flavors, which can mix, through Yukawa-induced running, into operators that contribute to $0\nu\beta\beta$ decays at tree level.

We have found that RG-induced constraints from $0\nu\beta\beta$ decays on operators of the type $\psi^4 H$ and $\psi^4 D$ with 2nd and 3rd generation quarks are, for most of the flavor combinations, stronger than the limits derived from meson decays, such as $\smash{P^+\to e^+ +\mathrm{inv}}$, $\smash{P^+\to P^{\prime-} e^+ e^+}$ and $\smash{P^+\to P^{\prime+} +\mathrm{inv}}$\,, where $\smash{P^{(\prime)}}$ stand for generic pseudoscalar mesons. With current experimental limits on $0\nu\beta\beta$ decays~\cite{KamLAND-Zen:2024eml}, we are able to probe an effective New Physics scale $\Lambda$ as large as to $30~\mathrm{TeV}$ through RG effects, depending on the type of operator and the specific quark flavor, cf.~Fig.~\ref{fig:eft-limits-quark}. The sensitivity to $\Lambda$ is expected to improve by about a factor of $2$ with the next-generation experiments~\cite{LEGEND:2021bnm,nEXO:2021ujk,SNO:2021xpa}.

The structure of our results becomes more transparent by expanding the leading-logarithmic solution of the RG equations in Eq.~\eqref{eq:rge-d7-solution}. While most of the dominant contributions arise through the first leading-logarithm, there are specific operators for which the leading effects appear only at higher orders in this expansion. Indeed, this is the case if the one-loop mixing between operators is suppressed, so that the leading effects arise through a two-step mixing. This observation highlights the phenomenological importance of determining the two-loop anomalous dimensions in a consistent renormalization scheme for fully accounting for these potential effects.

In addition to exploring the impact of two-loop RG contributions, another promising direction is to investigate the implications of our results for TeV-scale ultraviolet models of neutrino masses, which generate predominantly dimension-seven operators with heavy quarks. A complete phenomenological analysis must also account for numerous constraints from low- and high-energy observables on the new mediators. Such an investigation lies beyond the scope of the present work and will be the subject of future studies.

\section*{Acknowledgments}

This project has received support from the Agence
Nationale de la Recherche under the contract ANR25-CE31-2504, from the European Union’s Horizon 2020 research and innovation programme under the Marie Skłodowska-Curie grant agreement N$^\circ$~860881-HIDDeN and N$^\circ$~101086085-ASYMMETRY, from the IN2P3 (CNRS) Master Project HighPTflavor, from the SPRINT/CNRS agreement supported by FAPESP under Contract No.~2023/00643-0 and from the USP-COFECUB project Uc Ph194-2. S.F.~acknowledges the financial support from the Slovenian Research Agency (research core funding No.~P1-0035 and N1-0321). L.L.~acknowledges the financial support of FAPESP under the grant number 2021/02283-6.   R.Z.F.  thanks IJCLab for their hospitality during part of this work, as well as CNPq for partial financial support.
JdV is funded by the Dutch Research Council (NWO) in the form of a VIDI grant and by the European Union (ERC, CRUNS, 101230525). Views and opinions expressed are, however,  those of the author(s) only and do not necessarily reflect those of the European Union or the European Research Council. Neither the European Union nor the granting authority can be held responsible for them.
\appendix 

\section{Notation and conventions}
\label{app:notation}

In this Appendix, we briefly introduce our notation and conventions. The SM Lagrangian is expressed as 
\begin{align}
    \mathcal{L}_{\mathrm{SM}} &= -\dfrac{1}{4}G_{\mu\nu}^A G^{A\,\mu\nu} -\dfrac{1}{4}W_{\mu\nu}^I W^{I\,\mu\nu} -\dfrac{1}{4}B_{\mu\nu} B^{\mu\nu}\\*[0.3em]
    &+\big{(}D_\mu H^\dagger\big{)} \big{(}D_\mu H\big{)}-\mu_H^2\,H^\dagger H - \lambda \big{(}H^\dagger H\big{)}^2 \nonumber\\[0.3em]
    &+\sum_\psi \bar{\psi} i \slashed{D} \psi+ \mathcal{L}_{\mathrm{yuk}}\,, \nonumber
\end{align}
where $\psi\in \lbrace Q,L,u,d,e\rbrace$ and the covariant derivative is defined with the convention $D_\mu = \partial_\mu+i g_3 T^A G_\mu^A+i g_2 t^I W_\mu^I+i g_1 Y B_\mu$, where $T^A$ are the $SU(3)$ generators, $t^I$ are the $SU(2)$ generators and $Y$ is the $U(1)$ hypercharge generator. The Yukawa Lagrangian reads 
\begin{align}
\mathcal{L}_{\mathrm{yuk}}=-y_u\,\bar{Q} \tilde{H} u-y_d\,\bar{Q}{H}d-y_\ell\,\bar{L}{H}e + \mathrm{h.c.}\,,
\end{align}
where flavor indices are omitted.  We work in the flavor basis with diagonal down-type quark and lepton Yukawas. Quark and lepton flavor indices are denoted in this paper by Latin and Greek symbols, respectively.

\section{Matching at the electroweak scale}
\label{app:matching}

In this Appendix, we provide the tree-level matching of the LEFT defined in Eq.~\eqref{eq:left-d6} to the dimension-seven ones introduced in Eq.~\eqref{eq:smeft-d7}. For brevity, we define the shorthand notation $\overline{C} \equiv  C \,\Lambda^3/ v^3$ for the Wilson coefficients in the LEFT. The matching relations for dimension-six coefficients, at the electroweak scale $\mu \simeq \mu_{\mathrm{ew}}$, is given by 
\begin{align}
\label{eq:d6-matching-1}
\overline{C}_{\substack{V_L \\ ij}}^{(6)} &= -\dfrac{1}{\sqrt{2}}  V_{ij}\, {\mathcal{C}^{(7)\,\ast}_{\substack{LeHD\\ 11}}} + 4 V_{ij} \dfrac{m_{e}}{vg_2} \, {\mathcal{C}^{(7)\,\ast}_{\substack{LHW\\ 11}}}\,,\\*[0.55em]
\overline{C}_{\substack{V_R \\ ij}}^{(6)} &= \dfrac{1}{2\sqrt{2}} {\mathcal{C}^{(7)\,\ast}_{\substack{\bar{d}LueH\\j 1 i 1}}}\,, \nonumber\\*[0.55em]
\overline{C}_{\substack{S_L \\ ij}}^{(6)} &= {\dfrac{1}{\sqrt{2}}\,\mathcal{C}^{(7)\,\ast}_{\substack{\bar{Q}uLLH\\ji11}}}+\dfrac{V_{ij}}{2} \dfrac{m_{u_i}}{v} \mathcal{C}_{\substack{LHD2\\ 11}}^{(7)\,\ast}\,,  \nonumber\\*[0.55em]
\overline{C}_{\substack{S_R \\ ij}}^{(6)} &= \dfrac{1}{2\sqrt{2}}\sum_{k}V_{ik} \,\mathcal{C}_{\substack{\bar{d}LQLH1\\j 1 k 1}}^{(7)\,\ast} -\dfrac{V_{ij}}{2} \dfrac{m_{d_j}}{v}\mathcal{C}_{\substack{LHD2 \\11}}^{(7)\,\ast}\,,  \nonumber\\*[0.55em]
\overline{C}_{\substack{T \\ij}}^{(6)}   &= \dfrac{1}{8\sqrt{2}}\sum_{k}V_{ik} \Big{(}2\,\mathcal{C}^{(7)\,\ast}_{\substack{\bar{d}LQLH2\\j1k1}}+\mathcal{C}^{(7)\,\ast}_{\substack{\bar{d}LQLH1\\j1k1}}\Big{)}\,.  \nonumber
\end{align}
Similarly, for the relevant dimension-seven operators, we obtain
\begin{align}
\label{eq:d6-matching-2}
    \overline{C}_{\substack{V_L }}^{(7)}&= -\frac{V_{ud}}{2}\Big{(}2\,\mathcal{C}_{\substack{LHD1\\
    11}}^{(7)\,\ast} + \mathcal{C}_{\substack{LHD2 \\11}}^{\ast} \Big{)}+ \frac{8}{g_2}{\mathcal{C}^\ast_{\substack{LHW\\ 11}}}\,,\\*[0.35em]
    \overline{C}_{\substack{V_R }}^{(7)}&= -\mathcal{C}_{\substack{\bar{d}uLLD\\
    1111}}^{(7)\,\ast}\,, \nonumber
\end{align}

\noindent and for the dimension-nine ones,
\begin{align}
    \overline{C}_{\substack{1}}^{(9)} &= -2 \,V_{ud}^2 \Big{(}{\mathcal{C}^{(7)\,\ast}_{\substack{LHD1\\ 11 }}}+\frac{4}{g_2}{\mathcal{C}^{(7)\,\ast}_{\substack{LHW\\ 11 }}}\Big{)}\,,\\[0.4em]
    \overline{C}_{4}^{(9)} &= -2 \, V_{ud} \,{\mathcal{C}^{(7)\,\ast}_{\substack{\bar{d}uLLD\\ 1111 }}}\,. \nonumber 
\end{align}

\section{LNV meson decays}
\label{app:lnv-meson-formulas}

In this Appendix, we provide the expressions for the $P^-\to P^{\prime +} \ell^-\ell^-$ process, in the presence of the charged-current operators given in Eq.~\eqref{eq:left-d6}, as illustrated in panel \emph{c)} of Fig.~\ref{fig:diagrams-flavor}. We consider the diagrams with a single insertion of these operators, together with an insertion of the charged-current Fermi theory operator. We compute these contributions via the vacuum insertion approximation,~\footnote{Notice that connected topologies are also possible, which cannot be factorized in this way, cf.~Ref.~\cite{Donini:2025cuy}. We neglect these contributions, as they cannot be computed from first principles. See also Ref.~\cite{Zhou:2021lnl} for a complete calculation of $K^+\to\pi^-\ell^+\ell^+$ based on Chiral Perturbation Theory.}
\begin{equation}
    \label{eq:vip}
    \langle P^\prime | J_{kl} \times J_{ij} |P\rangle \sim \langle P^\prime | J_{kl} | 0 \rangle  \, \langle 0 | J_{ij} | P \rangle\,,
\end{equation}

\noindent where $J_{ij}$ and $J_{kl}$ are hadronic currents, and $P=\bar{u}_i d_j$ and $P^\prime=\bar{u}_k d_l$ are pseudoscalar mesons. The matrix elements on the right-hand side of Eq.~\eqref{eq:vip} can then be expressed in terms of the meson decay constants $f_P$ and $f_{P^\prime}$~\cite{FlavourLatticeAveragingGroupFLAG:2024oxs}. We can then decompose the $P^-\to P^{\prime +} \ell^-\ell^-$ differential decay rates as follows,
\begin{align}
    \dfrac{\mathrm{d}\Gamma}{\mathrm{d}q^2} &= \dfrac{G_F^4 f_P^2 f_{P^\prime}^2 \lambda_P^{1/2} } {2 (4\pi)^3 m_P^3}  \bigg{[} q^2\, |\mathcal{F}_S|^2 + \dfrac{\lambda_P}{12 m_P^2}\,|\mathcal{F}_V|^2 \bigg{]}\,,
\end{align}

\noindent where $\lambda_P \equiv \lambda (q^2, m_P^2, m_{P^\prime}^2)$, $q^2\equiv m_{\ell\ell}^2$ is the dilepton invariant-mass, $m_{P^{(\prime)}}$ denote the meson masses, and we have neglected the electron mass.  There are, in principle, two diagrams, with the insertion of the New Physics contribution at either the $u_i\to d_j e \nu$ or the $u_k\to d_l e \nu$ transition, which are described by the functions,

\begin{align}
\mathcal{F_S} &=\big{(}C_{\substack{S_L\\ij}}-C_{\substack{S_R\\ij}}\big{)}\dfrac{ m_P^2\,V_{kl}}{(m_{u_i}+m_{d_j})}+\big{(}C_{\substack{S_L\\kl}}-C_{\substack{S_R\\kl}}\big{)}\dfrac{ m_{P^\prime}^2\,V_{ij}}{(m_{u_k}+m_{d_l})}\,,\nonumber\\*[0.35em]
\mathcal{F}_V &=\big{(}C_{\substack{V_L\\ij}}-C_{\substack{V_R\\ij}}\big{)}\,V_{kl}+\big{(}C_{\substack{V_L\\kl}}-C_{\substack{V_R\\kl}}\big{)}\,V_{ij}\,.
\end{align}

\noindent where we consider contributions up to $\mathcal{O}(\Lambda^{-3})$ at the amplitude level.

\end{document}